\documentclass[10pt, twocolumn, final]{IEEEtran}
\usepackage{algorithm, algorithmic}
\usepackage{epsfig, graphics, color}
\usepackage{amsmath, amssymb, amsfonts, amsbsy, mathrsfs, bm}
\usepackage{cases}
\usepackage{enumerate, fancyvrb}
\usepackage{theorem}
\usepackage[compress]{cite}
\DeclareMathAlphabet{\mathsfsl}{OT1}{cmss}{m}{sl}

\ifCLASSOPTIONcompsoc
 \usepackage[tight, normalsize, sf, SF]{subfigure}
\else
 \usepackage[tight, footnotesize]{subfigure}
\fi

\title{Cooperative and Distributed Localization for Wireless Sensor Networks in Multipath Environments}
\author{Mei Leng, Wee Peng Tay, and Tony~Q.S.~Quek
\thanks{M. Leng and W.P. Tay are with the Nanyang Technological University, Singapore. (e-mail:$\{$lengmei,wptay$\}$@ntu.edu.sg).}
\thanks{T.Q.S. Quek is with the Institute for Infocomm Research, Singapore. (e-mail:qsquek@i2r.a-star.edu.sg).}}

\newcommand{\bP}{\mathbf{P}}

\newcommand{\bW}{\mathbf{W}}

\newcommand{\bs}{\mathbf{s}}

\newcommand{\bSigma}{\mathbf{\Sigma}}
\newcommand{\N}[2]{{\mathcal{N}\left(#1\ ;\ #2\right)}}
\newcommand{\tc}[1]{^{(#1)}}

\begin{document}
\maketitle

\begin{abstract}
We consider the problem of sensor localization in a wireless network in a multipath environment, where time and angle of arrival information are available at each sensor. We propose a distributed algorithm based on belief propagation, which allows sensors to cooperatively self-localize with respect to one single anchor in a multihop network. The algorithm has low overhead and is scalable. Simulations show that although the network is loopy, the proposed algorithm converges, and achieves good localization accuracy.
\end{abstract}


\section{Introduction}
A wireless sensor network (WSN) consists of many devices (or nodes) capable of onboard sensing, computing and communications. WSNs are used in industrial and commercial applications, such as environmental monitoring and pollution detection, control of industrial machines and home appliances, event detection, and object tracking \cite{Akyildiz2007,Bulusu2005,Tay2009}. In most applications, the data collected by the sensor nodes can only be meaningfully interpreted if it is correlated with the location of the corresponding sensors. The Global Positioning System (GPS) is widely used for localization in outdoor environments\cite{Bulusu2000}. However, GPS is a costly option and is not suitable for power-limited sensor nodes in WSNs. Furthermore, GPS signals do not penetrate well to indoor environments. Therefore, alternatives to GPS localization have been widely studied\cite{Bulusu2000, Wymeersch2009}. In a WSN, nodes whose positions are known are called ``\emph{anchors}''. By making use of pairwise range or angle measurements between anchors and/or other sensor nodes whose positions are unknown, sensors with no access to GPS can perform self-localization. Typical techniques include the use of time-of-arrival (TOA), time-difference-of-arrival (TDOA), received-signal-strength (RSS) and/or angle-of-arrival (AOA) information in triangulating the location of a node. This is usually studied in line-of-sight (LOS) environments\cite{Sun2005, Mao2007}.

However, LOS signals do not always exist in urban or cluttered environments, where signals usually experience multiple refections and diffractions. Such signals are referred to as nonline-of-sight (NLOS) signals and are commonly encountered in both indoor (e.g., residential buildings, offices and shopping malls) and outdoor (e.g., metropolitan and urban) environments. NLOS errors mitigation techniques have been extensively investigated \cite{Guvenc2009, Al-Jazzar2007, Cong2005, Al-Jazzar2009, Xie2009, Miao2007, Seow2008}, but most of the algorithms in the literature focus on locating one single sensor with several anchors. Since they require each sensor to have direct signal paths to anchors in the network, such algorithms cannot be applied in network-wide localization. On the other hand, several algorithms for network-wide localization have been proposed in the literature \cite{Biswas2006, Tseng2007, Srirangarajan2008, Wymeersch2009, Ihler2005}. Unfortunately, only LOS signals are considered in these algorithms.

Distributed localization algorithms for multipath environments were proposed in \cite{Ananthasubramaniam2008,Ekambaram2010}, where NLOS error is modeled as a positive bias in range and angle measurements, and its statistical characteristics are inferred by numerical methods, such as bootstrap sampling in \cite{Ananthasubramaniam2008}, and particle filters in \cite{Ekambaram2010}. One of the major disadvantages is that these Bayesian inference techniques require a large number of observations and are computationally expensive. Generally, the statistical model for NLOS errors depends on various factors, such as signal bandwidth, propagation medium and environment temperature. Different models including the uniform, exponential and Rayleigh distributions, have been proposed in the literature \cite{Pedersen2000,Zekavat2011}.

Instead of modeling NLOS errors in multipath environments as random biases, geometric analysis can be applied in pairwise localization \cite{Xie2009,Miao2007,Seow2008}, where measurements of different paths are modeled using closed-form expressions, which significantly simplifies the system model. In this paper, we derive a distributed localization algorithm based on range and direction measurements at each node, and where nodes exchange information to cooperatively perform self-localization relative to a fixed reference node. We show through simulation that by exchanging limited information, all the nodes in the network can perform localization to a good accuracy. We compare the performance to that achieved without cooperation, and show that cooperation amongst neighboring nodes significantly improves the localization accuracy.

The rest of this paper is organized as follows. In Section \ref{Section:System}, we define the system model. We describe our algorithm in Section \ref{Section:DistributedLocalization}, and provide simulation results in Section \ref{Section:SimulationResults}. In Section \ref{Section:Conclusions}, we summarize and conclude.

\section{System model}\label{Section:System}
Consider a network of $M+1$ sensors, $\{S_0, S_1, \cdots, S_M\}$. The position of $S_i$ is $\mathbf{s}_i \triangleq (x_i, y_i)$, where $x_i$ and $y_i$ are its $x$- and $y$-coordinates respectively. Without loss of generality, we assume that the position of $S_0$ is known and that $\mathbf{s}_0 = (0,0)$. The objective of each $S_i$ is to perform self-localization relative to $S_0$.

\begin{figure}[!t]
\centering
\includegraphics[width=0.35\textwidth]{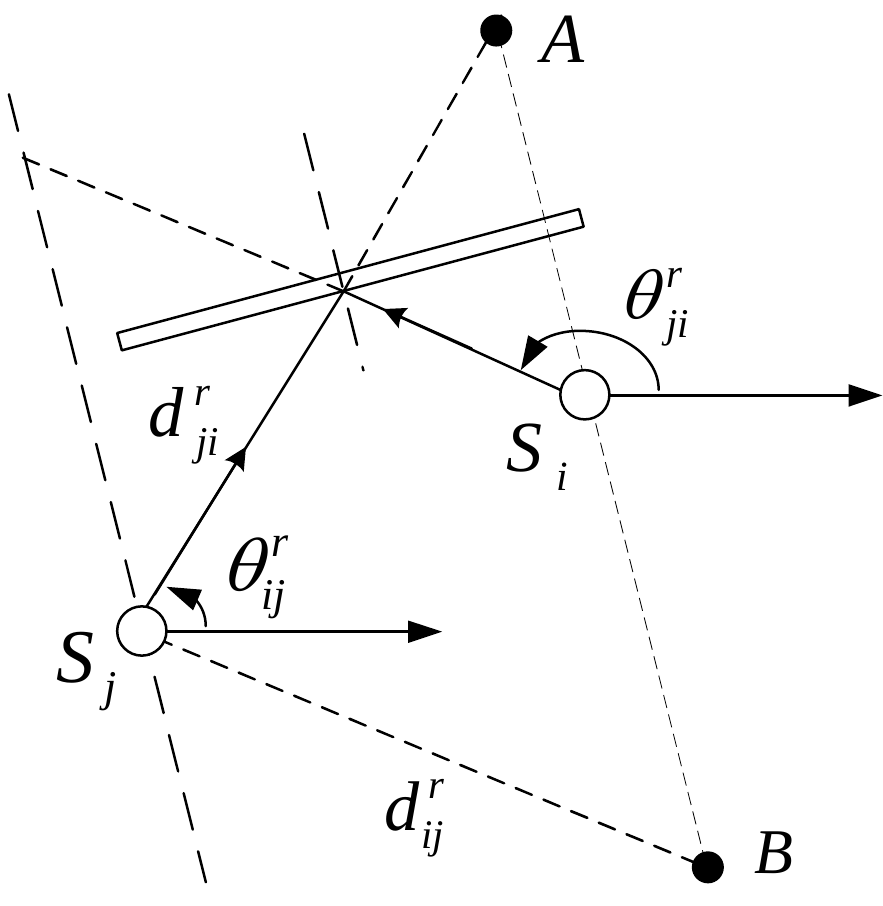}
\caption{An example for single-bounce scattering path between $S_i$ and $S_j$.} \label{Figure-Scatter}
\end{figure}

In the following, we consider two nodes $S_i$ and $S_j$. Similar to \cite{Seow2008}, we describe a model to relate the range and direction measurements at each node to their positions. Suppose that there are $R$ LOS or NLOS paths between $S_i$ and $S_j$. An example of a single-bounce scattering path is shown in Figure \ref{Figure-Scatter}, where the signal from $S_j$ to $S_i$ is reflected at a nearby scatter, and the communication link between two nodes is assumed to be symmetric. Let $d_{ji}^{r}$ be the distance measured by $S_i$ using the time-of-arrival information of the signal along the $r^{\textrm{th}}$ path from $S_j$, and $\theta_{ji}^r$ be the corresponding angle-of-arrival information. Nodes $S_i$ and $S_j$ exchange measurements with each other, so that both nodes have the measurements $\{d_{ij}^{r}, d_{ji}^{r}, \theta_{ij}^{r}, \theta_{ji}^{r}\}_{r = 1}^{R}$.

Consider the $r^{\textrm{th}}$ path between $S_i$ and $S_j$. Given the position of $S_j$ and $\{d_{ij}^{r}, d_{ji}^{r}, \theta_{ij}^{r}, \theta_{ji}^{r}\}$, the position of $S_i$ cannot be determined with certainty even in the absence of measurement and communication noise. As shown in Figure \ref{Figure-Scatter}, the estimated position for $S_i$ can be any point along the line $AB$. If there are multiple paths between $S_i$ and $S_j$ from non-parallel scatters, the position of $S_i$ can be found as the intersection point of two such lines. Suppose that there is no measurement noise, then a straightforward geometric consideration shows that
\begin{align}
\mathbf{p}_{A} - \mathbf{p}_{B} =
\begin{bmatrix}
d_{ji}^{r}\cos(\theta_{ij}^{r}) + d_{ij}^{r}\cos(\theta_{ji}^{r}) \\
d_{ji}^{r}\sin(\theta_{ij}^{r}) + d_{ij}^{r}\sin(\theta_{ji}^{r})
\end{bmatrix},
\end{align}
where $\mathbf{p}_{A}$ and $\mathbf{p}_{B}$ are the positions of $A$ and $B$ respectively. A vector perpendicular to $\mathbf{p}_A - \mathbf{p}_B$, is
\begin{align}
\mathbf{n}_{AB} =
\begin{bmatrix}
- d_{ji}^{r}\sin(\theta_{ij}^{r}) - d_{ij}^{r}\sin(\theta_{ji}^{r}) \\
d_{ji}^{r}\cos(\theta_{ij}^{r}) + d_{ij}^{r}\cos(\theta_{ji}^{r})
\end{bmatrix}.
\end{align}
Since both $A$ and $S_i$ are on the line $AB$, we have $\mathbf{n}_{AB}^{T} \mathbf{s}_i = \mathbf{n}_{AB}^{T} \mathbf{p}_A$, from which we obtain
\begin{align*}
& \begin{bmatrix}
- d_{ji}^{r}\sin(\theta_{ij}^{r}) - d_{ij}^{r}\sin(\theta_{ji}^{r}) \\
d_{ji}^{r}\cos(\theta_{ij}^{r}) + d_{ij}^{r}\cos(\theta_{ji}^{r})
\end{bmatrix}^{T} \mathbf{s}_i \\
& =
\begin{bmatrix}
- d_{ji}^{r}\sin(\theta_{ij}^{r}) - d_{ij}^{r}\sin(\theta_{ji}^{r}) \\
d_{ji}^{r}\cos(\theta_{ij}^{r}) + d_{ij}^{r}\cos(\theta_{ji}^{r})
\end{bmatrix}^{T} \left\{ \mathbf{s}_j + \begin{bmatrix}
                                                       d_{ji}^{r} \cos(\theta_{ij}^{r}) \\
                                                       d_{ji}^{r} \sin(\theta_{ij}^{r})
                                                      \end{bmatrix}\right\},
\end{align*}
which can be further simplified as
\begin{align}\label{relation}
d_{ji}^{r} & =
\mathbf{g}(\theta_{ij}^r, \theta_{ji}^r)^{T} \left(\mathbf{s}_i - \mathbf{s}_j\right),
\end{align}
where
\begin{align*}
\mathbf{g}(\theta_{ij}^r, \theta_{ji}^r) & =
\begin{bmatrix}
\frac{\sin(\theta_{ij}^{r}) + \sin(\theta_{ji}^{r})}{\sin(\theta_{ji}^{r} - \theta_{ij}^{r})} \\
-\frac{\cos(\theta_{ij}^{r}) + \cos(\theta_{ji}^{r})}{\sin(\theta_{ji}^{r} - \theta_{ij}^{r})}
\end{bmatrix}.
\end{align*}
We have made use of the fact that $d_{ij}^r = d_{ji}^r$ for symmetric communication links between $S_i$ and $S_j$ in \eqref{relation}. We have not factored in measurement and communication noise up to this point. Let the corresponding noisy measurements be $\{\tilde{d}_{ji}^r, \tilde{\theta}_{ji}^r, \tilde{\theta}_{ij}^r\}$. Modeling the total effect of noise as a Gaussian random error $\varpi_{ji}^r$, we have
\begin{align}
\tilde{d}_{ji}^r = \mathbf{g}\left(\tilde{\theta}_{ji}^r, \tilde{\theta}_{ij}^r\right)^{T}\left(\mathbf{s}_i - \mathbf{s}_j\right) + \varpi_{ji}^r.
\end{align}
We assume that measurements for the $R$ paths are such that $\left\{\mathbf{g}\left(\tilde{\theta}_{ji}^r, \tilde{\theta}_{ij}^r\right): r=1,\ldots,R\right\}$ are linearly independently, otherwise some of the paths are duplicates of each other. We also assume that the noise terms $\varpi_{ji}^r$ are i.i.d. Gaussian random variables with zero mean and variance $\sigma^2$. Stacking the measurements from all $R$ paths into a vector $\mathbf{d}_{ji} = [\tilde{d}_{ji}^1, \ldots, \tilde{d}_{ji}^R]^T$, and letting $\mathbf{G}_{ji} = \left[\mathbf{g}\left(\tilde{\theta}_{ji}^1, \tilde{\theta}_{ij}^1\right), \ldots, \mathbf{g}\left(\tilde{\theta}_{ji}^R, \tilde{\theta}_{ij}^R\right)\right]^T$, and $\bm{\varpi}_{ji} = [\varpi_{ji}^1,\ldots,\varpi_{ji}^R]^T$, we can estimate the position of $S_i$ from $\mathbf{s}_j$ using
\begin{align}
\mathbf{s}_i = \mathbf{s}_j + \mathbf{G}_{ji}^{\dagger}\left(\mathbf{d}_{ji} - \bm{\varpi}_{ji}\right), \label{si-sj}
\end{align}
where $\mathbf{G}_{ji}^{\dagger}$ is the Moore-Penrose pseudo-inverse of the matrix $\mathbf{G}_{ji}$. If there are more than one signal paths between nodes $S_i$ and $S_j$, we have $\mathbf{G}_{ji}^{\dagger} = \left(\mathbf{G}_{ji}^T\mathbf{G}_{ji}\right)^{-1}\mathbf{G}_{ji}^T$. If there is only one signal path, $\mathbf{G}_{ji}^{\dagger} = \mathbf{G}_{ji}^T\left(\mathbf{G}_{ji}\mathbf{G}_{ji}^T\right)^{-1}$. Let $\bSigma_{ji} = \mathbf{G}_{ji}^{\dagger} (\mathbf{G}_{ji}^{\dagger})^T$. The posterior distribution of the node locations is given by
\begin{align*}
p(\mathbf{s}_i - \mathbf{s}_j \mid \mathbf{G}_{ji}^{\dagger}\mathbf{d}_{ji}, \mathbf{G}_{ji})
& = \N{\mathbf{s}_i - \mathbf{s}_j}{\mathbf{G}_{ji}^{\dagger}\mathbf{d}_{ji}, \sigma^2\bSigma_{ji}}.
\end{align*}

Similar analysis in \cite{Seow2008} proposes a least square estimator to localize a single node with respect to a reference node. However, to localize every node in a network, it is necessary to consider the interaction between $S_i$ and all the other nodes $\{S_j\}_{j = 0, j \neq i}^{M}$. The MAP estimator for $\mathbf{s}_i$ is given by (\ref{joint}) on top of next page.
\begin{figure*}[!t]
\setcounter{equation}{6}
\begin{equation}\label{joint}
\hat{\mathbf{s}}_i = \operatorname*{arg\,max}_{\mathbf{s}_i} \ \operatorname*{\int\cdots\int}_{\mathbf{s}_0, \cdots, \mathbf{s}_{i-1}, \mathbf{s}_{i+1}, \cdots, \mathbf{s}_{M}} p\left(\mathbf{s}_i, \{\mathbf{s}_j\}_{j=0, j \neq i}^{M} \bigg\vert \left\{\mathbf{G}_{ji}^{\dagger}\mathbf{d}_{ji}, \mathbf{G}_{ji}\right\}_{j = 0, j \neq i}^{M}\right) \mathrm{d} \mathbf{s}_0 \cdots \mathrm{d} \mathbf{s}_{i-1} \mathrm{d} \mathbf{s}_{i+1} \cdots \mathrm{d} \mathbf{s}_M.
\end{equation}
\hrulefill
\end{figure*}
The joint posterior distribution in (\ref{joint}) depends on interactions amongst all the variables and is difficult to compute by brute-force integration.

In order to provide a computationally efficient algorithm to calculate the marginal distributions, we observe that the state at any sensor depends directly only on its neighboring sensors whose number is usually far less than the total number of variables. To explore such conditional independence structure, we make use of belief propagation in the following and propose an efficient algorithm which computes a set of marginal distributions from the joint posterior distribution without performing a full integration as in (\ref{joint}).

\emph{Remark 1}: When there exists a LOS path between $S_i$ and $S_j$, it is easy to see that $\vert \theta_{ji}^r - \theta_{ij}^r \vert  = \pi$ and $d_{ji}^{r} = \vert \mathbf{s}_i - \mathbf{s}_j \vert$, which is a special case of (\ref{relation}).

\emph{Remark 2}: When there exists paths with multiple bounces between $S_i$ and $S_j$, a two-step proximity detection scheme suggested in \cite{Seow2008} can be applied to detect and discard such paths, leaving measurements from either LOS and/or single-bounce paths for processing.

\section{Distributed localization based on belief propagation}\label{Section:DistributedLocalization}
\begin{figure}[!t]
\centering
\includegraphics[width=0.35\textwidth]{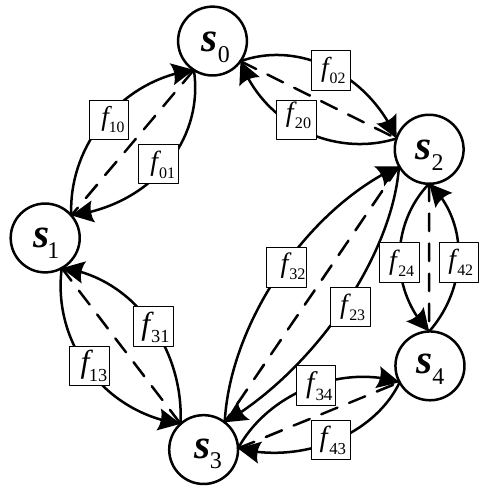}
\caption{An example for the factor graph of a network with $5$ nodes, where dashed lines indicate that a one-bounce scattering path exists between the corresponding two nodes, and the arrows indicate the direction of message flows.} \label{Figure-FG}
\end{figure}

We use Belief Propagation (BP) on a factor graph\cite{Kschischang2001} in this paper. An example for a network with $5$ nodes is shown in Figure \ref{Figure-FG}. Each random variable $\mathbf{s}_i$ is represented by a variable node (circle). The interaction between two sensors $S_i$ and $S_j$ is represented by a factor node (square) connected to both variable nodes $\mathbf{s}_i$ and $\mathbf{s}_j$. We split the interaction between $S_i$ and $S_j$ into two factors $f_{ji} \triangleq p(\mathbf{s}_i - \mathbf{s}_j \mid \mathbf{G}_{ji}^{\dagger}\mathbf{d}_{ji}, \mathbf{G}_{ji})$, and $f_{ij} \triangleq p(\mathbf{s}_j - \mathbf{s}_i \mid \mathbf{G}_{ij}^{\dagger}\mathbf{d}_{ij}, \mathbf{G}_{ij})$. Since messages only flow in one direction along the edges in the factor graph, they can be broadcast by the sensor nodes.

Without loss of generality, suppose the sensor $S_0$ is the anchor with a known position $(0, 0)$. For each variable $\mathbf{s}_i$, $i \in \{1, \cdots, M\}$, the marginal posterior distribution $m(\mathbf{s}_i)$ is found by iterative belief propagation, where two kinds of messages are involved,
\begin{itemize}
\item{$\mathit{b}^{(l)}_i(\mathbf{s}_i)$: belief of its own state at the variable node $\mathbf{s}_i$ after the $l^{\mathrm{th}}$ iteration,
\begin{align}
\setcounter{equation}{7}
\mathit{b}^{(l)}_i(\mathbf{s}_i) = \prod_{j \in \mathcal{B}_i}{\mathit{h}^{(l)}_{f_{ji} \rightarrow \bm{s}_i}(\mathbf{s}_i)}, \label{belief}
\end{align}
where $\mathcal{B}_i$ is the index set of $S_i$'s neighboring sensors.}
\item{$\mathit{h}^{(l)}_{f_{ji} \rightarrow \mathbf{s}_i}(\mathbf{s}_i)$: message from the factor node $f_{ji}$ to the variable node $\mathbf{s}_i$ in the $l^{\mathrm{th}}$ iteration, which represents $f_{ji}$'s belief of $\mathbf{s}_i$'s state, resulting from interactions between $\mathbf{s}_i$ and $\mathbf{s}_j$,
\begin{align}
\hspace{-1cm} \mathit{h}^{(l)}_{f_{ji} \rightarrow \mathbf{s}_i}(\mathbf{s}_i) = \int p(\mathbf{s}_i - \mathbf{s}_j \mid \mathbf{G}_{ji}^{\dagger}\mathbf{d}_{ji}, \mathbf{G}_{ji}) b^{(l-1)}_j(\mathbf{s}_j) \ \mathrm{d} \mathbf{s}_j. \label{message}
\end{align}}
\end{itemize}
Therefore, setting the initial belief $\mathit{b}^{(0)}_i(\mathbf{s}_i)$ to be the corresponding prior distribution $p(\mathbf{s}_i)$, beliefs (\ref{belief}) and messages (\ref{message}) are iteratively updated at each sensor, and the estimation for $\mathbf{s}_i$ is found by maximizing the converged belief $b^{(l)}_i(\mathbf{s}_i)$ with respect to $\mathbf{s}_i$. When prior distributions $\{p(\mathbf{s}_i)\}_{i=1}^{M}$ are Gaussian, closed-form expressions for beliefs and messages can be obtained as follows.

\subsection{Derivation for closed-form beliefs and messages}\label{Subsection:Derivation}
To derive closed-form expressions for (\ref{belief}) and (\ref{message}), we first consider the message from the anchor $S_0$ to a neighboring sensor $S_i$. Since $\mathbf{s}_0 \equiv (0, 0)$, its belief is a constant and can be represented as $b_0^{(l)}(\mathbf{s}_0) = \delta\{\mathbf{s}_0, (0, 0)\}$, where $\delta\{\cdot, \cdot\}$ is the Krocnecker delta function. Therefore, the message from $S_0$ to $S_i$ will be constant over all iterations, and is given by
\begin{align}
h_{f_{0i} \rightarrow \mathbf{s}_i}^{(l)}(\mathbf{s}_i) & = \int p(\mathbf{s}_i - \mathbf{s}_0 \mid \mathbf{G}_{0i}^{\dagger}\mathbf{d}_{0i}, \mathbf{G}_{0i}) \delta\{\mathbf{s}_0, (0,0)\} \ \mathrm{d} \mathbf{s}_0 \nonumber \\
& \propto \N{\mathbf{s}_i}{\bm{\nu}_{0i},\bW_{0i}}. \label{S0Si}
\end{align}
where $\bm{\nu}_{0i} = \mathbf{G}_{0i}^{\dagger}\mathbf{d}_{0i}$ and $\bW_{0i} = \sigma^2\bSigma_{0i}$ with $\bSigma_{0i} = \mathbf{G}_{0i}^{\dagger} (\mathbf{G}_{0i}^{\dagger})^T$.

Other the other hand, for all nodes $S_j$ other than $S_0$, we set the initial belief of $S_j$ as a Gaussian distribution with zero mean and large variance, i.e., for $j = 1, \cdots, M$, $b_j^{(0)}(\mathbf{s}_j) = \N{\mathbf{s}_j}{\bm{\mu}_j\tc{0}, \bP_j\tc{0}}$, where $\bm{\mu}_j\tc{0} = (0,0)$, and we define
\begin{align*}
\bP_j\tc{0} & = \left\{
\begin{array}{rl}
\alpha \mathbf{I}_2 & \text{if } |\mathcal{B}_j| > 1, \\
\frac{3\alpha}{2}\mathbf{I}_2 & \text{otherwise,}
\end{array}
\right.
\end{align*}
where $\alpha$ is a large positive value of at least an order of magnitude larger than the dimension of the environment in which the sensor nodes are located. Let $\lambda_{ij}^{\max}$ be the largest eigenvalue of $\bSigma_{ij}$. We assume that $\alpha \geq 2\sigma^2 \max_{i,j} \lambda_{ij}^{\max}$.

As the BP algorithm proceeds, the belief $b_j^{(l-1)}(\mathbf{s}_j)$ for $S_j$ after $l-1$ iterations is updated as $b_j^{(l-1)}(\mathbf{s}_j) = \N{\mathbf{s}_j}{\bm{\mu}_j^{(l-1)}, \mathbf{P}_j^{(l-1)}}$ and its message to $S_i$ in the $l^{\mathrm{th}}$ iteration is
\begin{align*}
h_{f_{ji} \rightarrow i}^{(l)}(\mathbf{s}_i) & = \int p(\mathbf{s}_i - \mathbf{s}_j \mid \mathbf{G}_{ji}^{\dagger}\mathbf{d}_{ji}, \mathbf{G}_{ji}) b_j^{(l-1)}(\mathbf{s}_j) \ \mathrm{d} \mathbf{s}_j \\
& \propto \N{\mathbf{s}_i}{\bm{\nu}_{ji}^{(l-1)}, \bW\tc{l-1}_{ji}},
\end{align*}
where
\begin{align}
\bm{\nu}_{ji}^{(l-1)}
& = \bm{\mu}_j^{(l-1)} + \mathbf{G}_{ji}^{\dagger}\mathbf{d}_{ji},\label{invmu} \\
\bW\tc{l-1}_{ji}
& = \sigma^2 \bSigma_{ji} + \mathbf{P}_j^{(l-1)}. \label{inv}
\end{align}
The message $h_{f_{ji} \rightarrow i}^{(l)}(\mathbf{s}_i)$ is a Gaussian distribution, so only the mean $\bm{\nu}_{ji}^{(l)}$ and covariance matrix $\bW\tc{l}_{ji}$ need to be passed to node $S_i$.

The belief of $\mathbf{s}_i$ after the $l^{\mathrm{th}}$ iteration then follows from (\ref{belief}). Since all the messages in the product in \eqref{belief} are Gaussian distributions, we have $b_i^{(l)}(\bm{s}_i) = \N{\bm{s}_i}{\bm{\mu}_i^{(l)}, \mathbf{P}_i^{(l)}}$ with
\begin{align}
\left[\mathbf{P}_i^{(l)}\right]^{-1} & = \sum_{j \in \mathcal{B}_i} \left[\bW\tc{l-1}_{ji}\right]^{-1} \label{P}
\end{align}
and
\begin{align}
\bm{\mu}_i^{(l)} = \mathbf{P}_i^{(l)} \sum_{j \in \mathcal{B}_i} \left[\bW\tc{l-1}_{ji}\right]^{-1} \bm{\nu}_{ji}^{(l-1)}. \label{mu}
\end{align}
At the end of the $l^{\mathrm{th}}$ iteration, sensor $S_i$ estimates its position by maximizing the belief $b^{(l)}_i(\mathbf{s}_i)$ with respect to $\mathbf{s}_i$. Since $b^{(l)}_i(\mathbf{s}_i)$ is a Gaussian distribution, the estimator for $\mathbf{s}_i$ at the $l^{\mathrm{th}}$ iteration is given by $\hat{\mathbf{s}}^{(l)}_i = \bm{\mu}^{(l)}_i$. This iterative procedure is formally given in Algorithm \ref{algorithm}. Notice that factor nodes are virtual, and are introduced only to facilitate the derivation of the algorithm. In practice, the update is performed at each individual sensor directly.

\begin{algorithm}[!t]
\caption{Distributed Localization in Multi-path Environments} \label{algorithm}
\begin{algorithmic}[1]
\STATE{\textbf{Initialization}:}
\STATE{Set the position at the anchor $S_0$ as $\mathbf{s}_0 = (0,0)$.}
\STATE{Set $\bm{\mu}_i^{(0)} = (0,0) $ and
\begin{align*}
\bP_i\tc{0} & = \left\{
\begin{array}{rl}
\alpha \mathbf{I}_2 & \text{if } |\mathcal{B}_i| > 1, \\
\frac{3\alpha}{2}\mathbf{I}_2 & \text{otherwise,}
\end{array}
\right.
\end{align*}}
\STATE{\textbf{Iteration until convergence}:}
\FOR{\textrm{the $l^{\textrm{th}}$ iteration}}
\STATE{\textbf{sensors} $S_i$ with $i = 1 : M$} \textbf{in parallel}
\STATE{broadcast the current belief $b^{(l-1)}_i(\mathbf{s}_i)$ to neighboring sensors;}
\STATE{receive $b^{(l-1)}_j(\mathbf{s}_j)$ from neighboring sensors $S_j$, where $j \in \mathcal{B}_i$;}
\STATE{update its belief as $b^{(l)}_i(\mathbf{s}_i) \sim \mathcal{N}(\bm{\mu}^{(l)}_i, \mathbf{P}^{(l)}_i)$ with
\begin{align*}
\left[\mathbf{P}_i^{(l)}\right]^{-1} = \sum_{j \in \mathcal{B}_i} \left[\bW\tc{l-1}_{ji}\right]^{-1},
\end{align*}
and
\begin{align*}
\bm{\mu}_i^{(l)} = \mathbf{P}_i^{(l)} \sum_{j \in \mathcal{B}_i} \left[\bW\tc{l-1}_{ji}\right]^{-1} \bm{\nu}_{ji}^{(l-1)},
\end{align*}
where $\bm{\nu}_{ji}^{(l-1)}$ and $\bW\tc{l-1}_{ji}$ are given in (\ref{inv}) and (\ref{invmu}) respectively. }
\STATE{estimate its position as $\hat{\mathbf{s}}_i^{(l)} = \bm{\mu}^{(l)}_i$.}
\STATE{\textbf{end parallel}}
\ENDFOR
\end{algorithmic}
\end{algorithm}

\section{Simulation Results}\label{Section:SimulationResults}
Numerical simulations are conducted to validate the effectiveness of our proposed algorithm. We consider a network with $5$ nodes randomly distributed in a $10\mathrm{m} \times 10\mathrm{m}$ square area. The factor graph is that in Figure \ref{Figure-FG}, where $\bs_0$ represents the anchor with a fixed location at $(0, 0)$, while the other nodes are the remaining sensors' locations. We set $\bs_1=(-4.5, -1.5)$, $\bs_2=(4.0,-1.0)$, $\bs_3=(-1.0,-8.0)$, and $\bs_4=(4.2,-6.0)$. Any two nodes that can communicate with each other through single-bounce scattering paths are connected with their connection indicated by a dashed line. The ranging measurement errors are i.i.d. Gaussian random variables with zero mean and standard variance $3$. The measurement error for AOA is assumed to be uniformly distributed in $[-5^{\circ}, 5^{\circ}]$. Each point in the figures is an average of $10^{4}$ independent simulation runs.

\begin{figure}
  \centering
  \subfigure[absolute errors on x-coordinates.]{
    \label{Figure-CDF-x-Ortho}
    \includegraphics[width=0.5\textwidth]{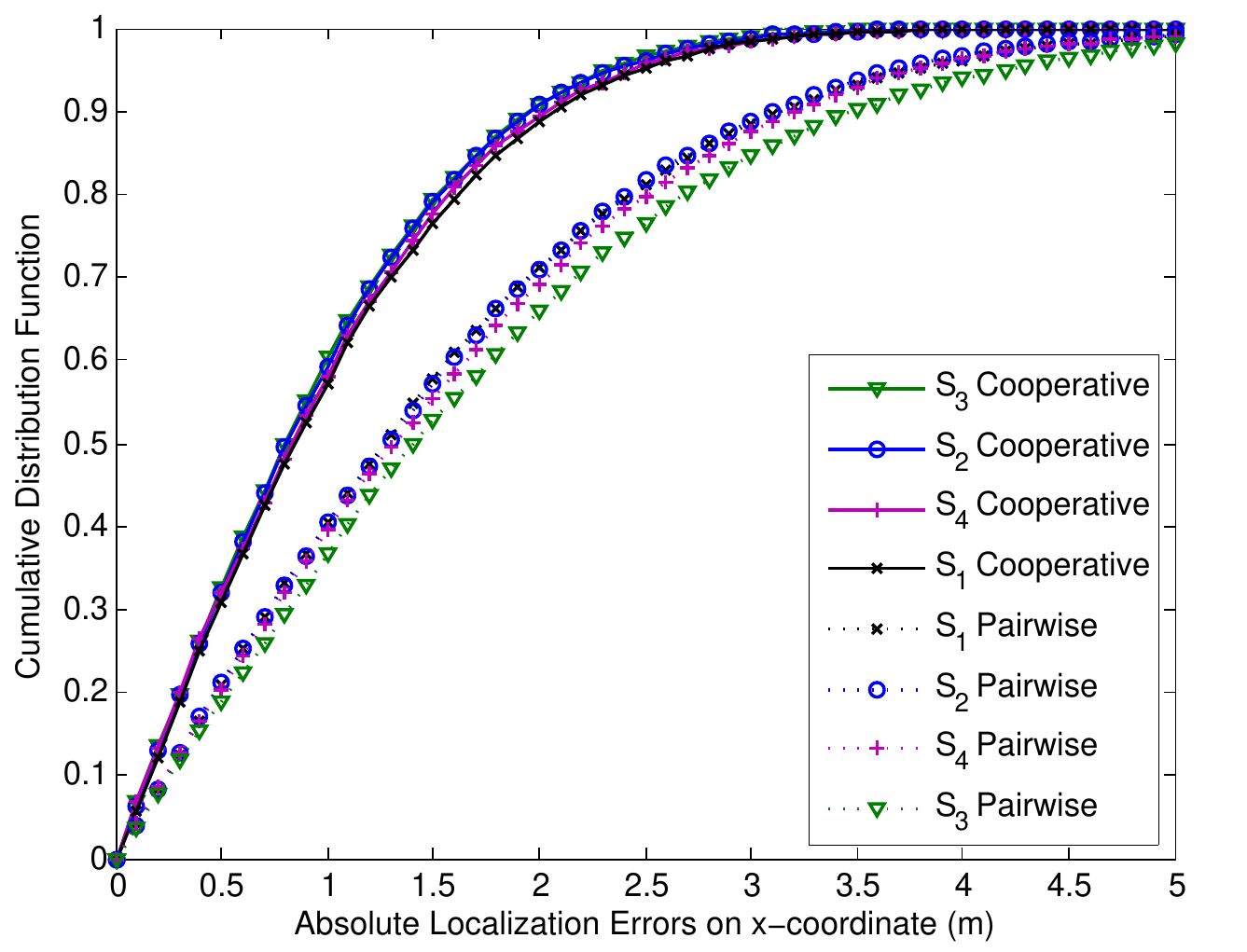}}
  \subfigure[absolute errors on y-coordinates.]{
    \label{Figure-CDF-y-Ortho}
    \includegraphics[width=0.5\textwidth]{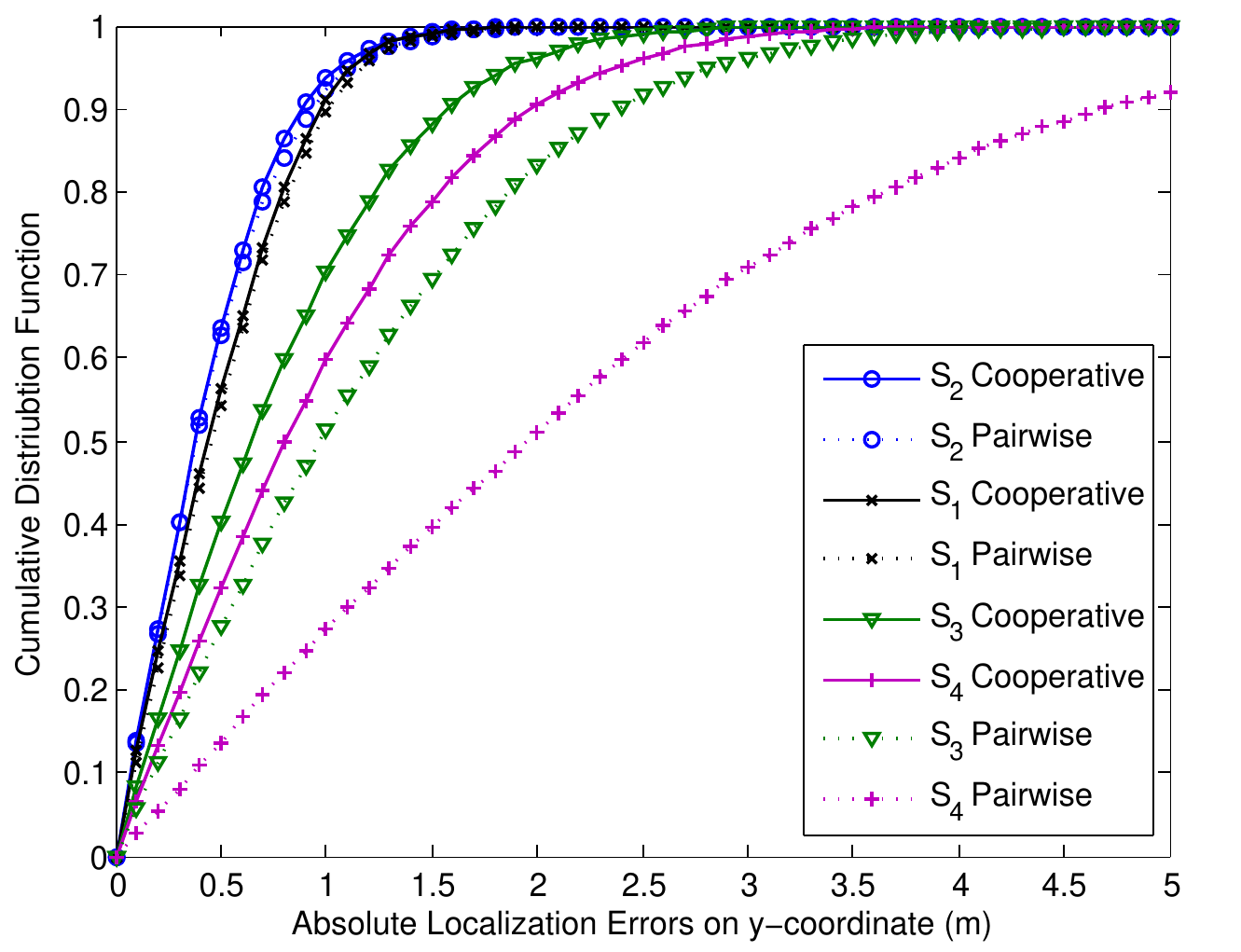}}
  \caption{Cumulative distribution function of absolute localization errors where scatters are orthogonal.}\label{Figure-CDF-Ortho}
\end{figure}

\begin{table*}[!t]
\caption[short title]{Mean error of estimated location at each sensor} \label{Table-ME}
\centering
\begin{tabular}{|c||c|c|c|c|c|c|}

\hline Localization scheme & Mean Error & $S_1$ & $S_2$ & $S_3$ & $S_4$ \\
\hline

Cooperative & $|\hat{x}_i - x_i|$ & $1.0048 \ \mathrm{m}$  &  $0.9558 \ \mathrm{m}$  &  $0.9459 \ \mathrm{m}$  &  $0.9834 \ \mathrm{m}$ \\
\cline{2-6}
Localization & $|\hat{y}_i - y_i|$ & $0.4914 \ \mathrm{m}$ &  $0.4385 \ \mathrm{m}$ &  $0.7639 \ \mathrm{m}$  & $0.9590 \ \mathrm{m}$ \\
\hline

Pairwise & $|\hat{x}_i - x_i|$ & $1.5119 \ \mathrm{m}$  &  $1.5034 \ \mathrm{m}$  &  $1.6823 \ \mathrm{m}$  &  $1.5529 \ \mathrm{m}$ \\
\cline{2-6}
Localization & $|\hat{y}_i - y_i|$ & $0.5145 \ \mathrm{m}$ &  $0.4544 \ \mathrm{m}$ &  $1.1562 \ \mathrm{m}$  & $2.2827 \ \mathrm{m}$ \\
\hline

\end{tabular} \\[3ex]
\end{table*}

First, we consider scenarios where scatters are orthogonal. We compare the performance of cooperative and pairwise localization. The cumulative distribution functions for absolute localization error of each sensor are shown in Figure \ref{Figure-CDF-Ortho}. Corresponding to the factor graph in Figure \ref{Figure-FG}, sensors $S_1$ and $S_2$ are directly connected to the anchor. Sensors $S_3$ and $S_4$ do not have any paths to the anchor. Instead each has a NLOS path to $S_1$ or $S_2$ respectively, and a NLOS path between themselves. In pairwise localization, $S_3$ localizes using only measurements from $S_1$, and $S_4$ localizes with respect to $S_2$. It can be seen from Figure \ref{Figure-CDF-x-Ortho} that $S_3$ and $S_4$ are localized with larger errors than $S_1$ and $S_2$, and this is because errors are accumulated over hops. In cooperative localization, $S_3$ and $S_4$ exchange information and incorporate measurements from the NLOS path between themselves. As shown in Figure \ref{Figure-CDF-x-Ortho}, the cooperative localization achieves better performances with more than $90\%$ of the localization errors less than $2 \mathrm{m}$ and all errors smaller than $3 \mathrm{m}$. Moreover, results for estimation on $y$-coordinates are shown in Figure \ref{Figure-CDF-y-Ortho}. It can be seen that both schemes have similar performances for sensors directly connected to the anchor (e.g., $S_1$ and $S_2$), and cooperation localization for $S_3$ and $S_4$ achieves better performances. The mean absolute errors for both schemes are further shown in Table \ref{Table-ME}.

\begin{figure}
  \centering
  \subfigure[absolute errors on x-coordinates.]{
    \label{Figure-CDF-x-Bitho}
    \includegraphics[width=0.5\textwidth]{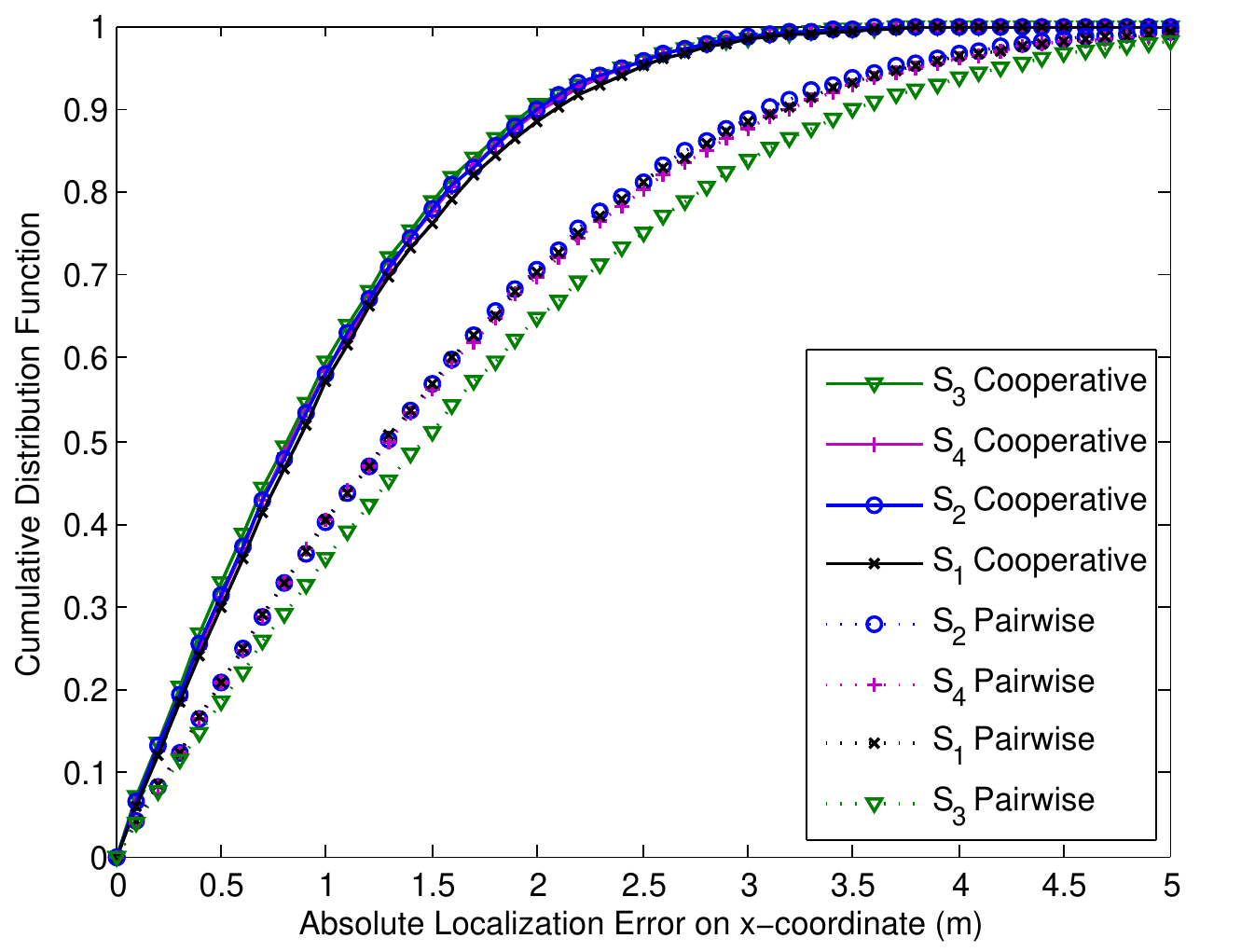}}
  \subfigure[absolute errors on y-coordinates.]{
    \label{Figure-CDF-y-Bitho}
    \includegraphics[width=0.5\textwidth]{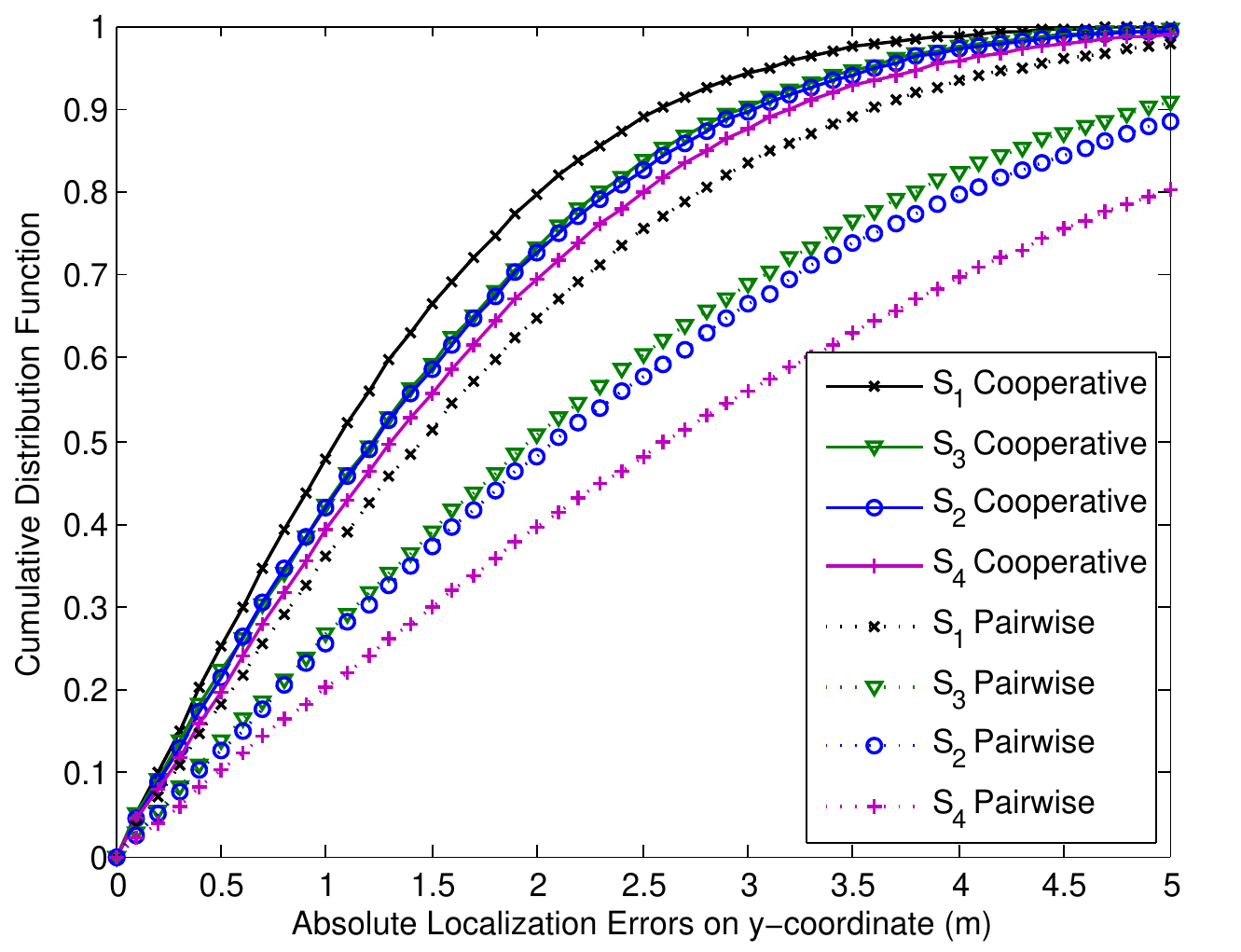}}
  \caption{Cumulative distribution function of absolute localization errors when scatters are horizontal or at $45^{\circ}$.} \label{Figure-CDF-Bitho}
\end{figure}

\begin{figure}[!t]
\centering
\includegraphics[width=0.45\textwidth]{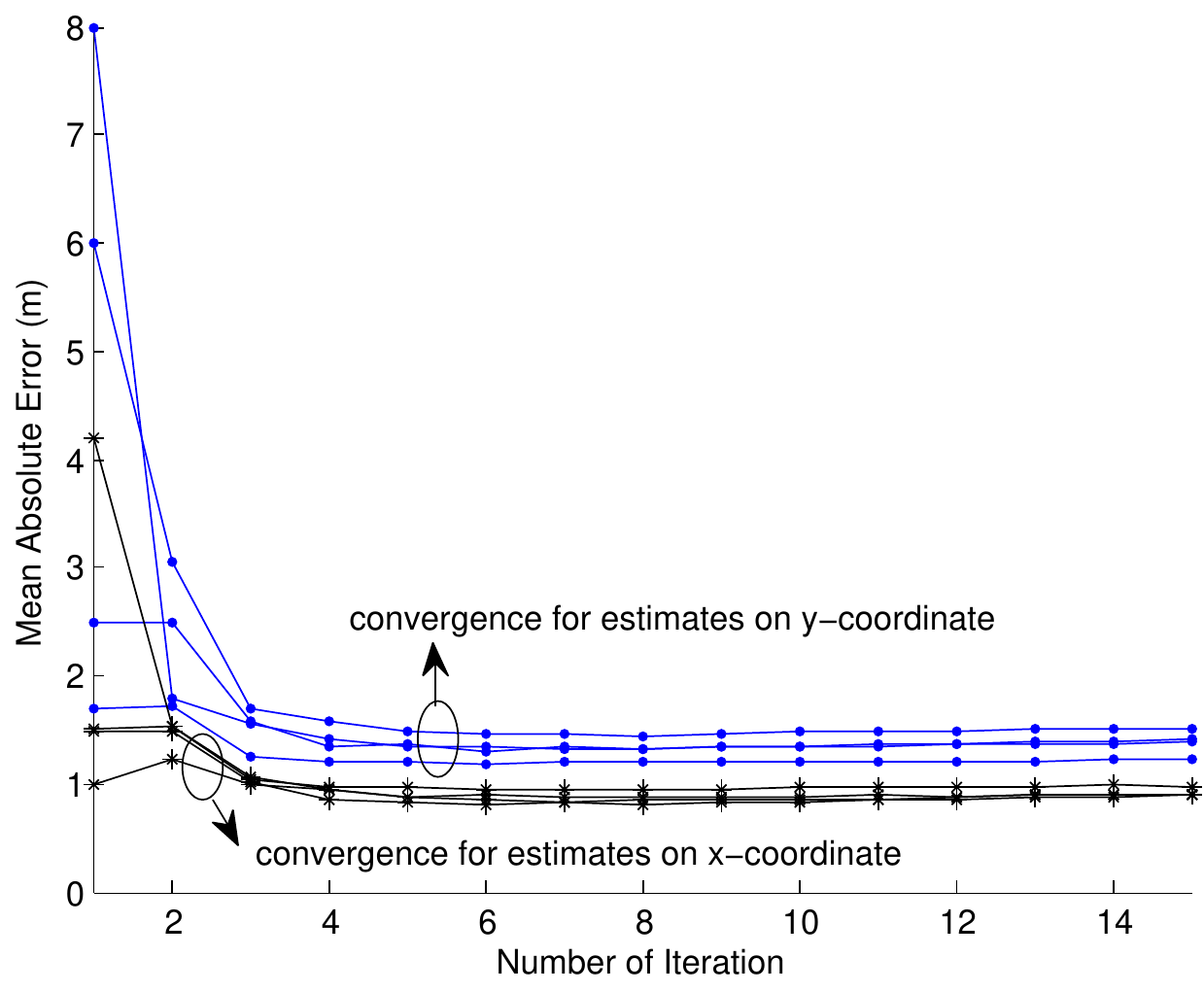}
\caption{Convergence of the mean absolute error when scatters are horizontal or at $45^{\circ}$, corresponding to that in Figure \ref{Figure-CDF-Bitho}.} \label{Figure-Convergence}
\end{figure}

Second, we consider scenarios where scatters are horizontal or at angle $45^{\circ}$ to the horizontal. As can be seen from Figure \ref{Figure-CDF-Bitho}, cooperation among neighboring sensors improves performance on both $x$- and $y$- coordinates. We also note that compared with Figure \ref{Figure-CDF-y-Ortho}, estimation errors for $y$-coordinates deteriorate due to correlation between measurements on the vertical direction. On the other hand, we investigate the convergence rate for our proposed algorithm in this biorthogonal scenarios, and the result is shown in Figure \ref{Figure-Convergence}. Simulations are also conducted when scatters are at $10^{\circ}$, $20^{\circ}$ and $30^{\circ}$ to the horizontal. Similar results as in Figure \ref{Figure-Convergence} are obtained and hence omitted here. These numerical results suggest that even for scenarios with non-orthogonal scatters, the mean of belief at each sensor converges in the proposed algorithm.

\section{Conclusions}\label{Section:Conclusions}
In this paper, we propose a distributed algorithm based on belief propagation for network-wide localization in multipath environments. The proposed algorithm requires communications only between neighboring sensors, with each sensor processing only information local to itself. The proposed algorithm has low overhead and can achieve robust and scalable localization. By utilizing both TOA and AOA information of the single-bounce scattering paths, we require only one anchor in the whole network, and sensors that do not have LOS/NLOS paths to the anchor can be localized by cooperation with its neighboring sensors. Simulation results show that our proposed algorithm achieve accuracy less than $1\mathrm{m}$ in a $10\mathrm{m} \times 10\mathrm{m}$ square area.

\bibliographystyle{IEEEtran}
\bibliography{IEEEabrv,Distributed_Localization}

\end{document}